\documentstyle{article}

\catcode`@=11

\long\def\@makefntext#1{\parindent 1em\noindent
 \hbox {$^{\@thefnmark})$\kern.5em}#1}
\renewcommand{\fnum@figure}{{\small\bf Fig.\kern.2em\thefigure\kern.1em}}
\renewcommand{\fnum@table}{{\small\bf Tab.\kern.2em\thetable\kern.1em}}

\catcode`@=12

\newcommand{\2}{\mbox{I{\kern-.1em}I}}
\newcommand{\3}{\mbox{\2{\kern-.1em}I}}

\newfam\mind
\textfont\mind=\textfont1
\scriptfont\mind=\scriptfont1
\scriptscriptfont\mind=\scriptscriptfont1
\def\mit{\fam\mind }

\newcommand{\ri}{\scriptfont1=\scriptfont0          %
                  \scriptscriptfont1=\scriptscriptfont\mind}
\newcommand{\ii}{\scriptfont1=\scriptfont\mind      %
                 \scriptscriptfont1=\scriptscriptfont\mind}
\newcommand{\rr}{\scriptfont1=\scriptfont0          %
                 \scriptscriptfont1=\scriptscriptfont0}

\newcommand{\rn}{\scriptfont1=\scriptfont0            %
                 \scriptscriptfont1=\scriptfont\mind} %

\newcommand{\beq}{\begin{equation}\ii}
\newcommand{\eeq}{\end{equation}}
\newcommand{\beqar}{\begin{eqnarray}\ii}
\newcommand{\eeqar}{\end{eqnarray}}

\newcommand{\SS}{\scriptstyle}
\newcommand{\TS}{\textstyle}
\newcommand{\DS}{\displaystyle}

\newcommand{\eq}[1]{\mbox{Eq.\kern.2em(\ref{#1})}}
\newcommand{\Eq}[1]{Equation~(\ref{#1})}
\newcommand{\fig}[1]{\mbox{Fig.\kern.2em\ref{fig_#1}}}
\newcommand{\Fig}[1]{Figure~\ref{fig_#1}}
\newcommand{\tab}[1]{\mbox{Tab.\kern.2em\ref{tab_#1}}}

\newcommand{\sect}[1]{\mbox{Sect.\kern.2em#1}}
\newcommand{\Sect}[1]{Section~#1}

\newcommand{\gp}{\gamma_{\rm p}}
\newcommand{\Ep}{E_{\rm p}}
\newcommand{\exval}[1]{\langle#1\rangle}
\newcommand{\D}[2]{\frac{\mbox{d}#1}{\mbox{d}#2}}
\renewcommand{\d}{\,\mbox{d}}
\newcommand{\FRII}{\mbox{FR-\2 }}

\newcommand{\ea}{et~al{}}

\newcommand{\lsim}{\mathrel{\raise.25ex\hbox{$<$}\kern-.8em\lower.9ex\hbox{$
\sim$}}}
\newcommand{\gsim}{\mathrel{\raise.25ex\hbox{$>$}\kern-.8em\lower.9ex\hbox{$
\sim$}}}

\let\ga = \gsim

\begin{document}

\title{Extragalactic ultra-high energy cosmic rays \\
\Large I. Contribution from hot spots in \FRII radio galaxies}

\author{J\"org P. Rachen and Peter L. Biermann \\
Max-Planck-Institut f\"ur Radioastronomie \\ Auf dem H\"ugel 69 \\
D-5300 Bonn 1, Germany}

\date{Accepted by Astronomy \& Astrophysics: December 30, 1992}

\maketitle

\ri      %

\begin{abstract}
The hot spots of Fanaroff-Riley class \2 radio galaxies, considered as working
surfaces of highly collimated plasma jets, are proposed to be the dominant
sources of the cosmic rays at energies above $1\,$EeV\footnotemark[1]. We
apply the model of first order Fermi acceleration at strong, nonrelativistic
shock waves to the hot spot region. The strength of the model has been
demonstrated by Biermann \& Strittmatter (1987) and by Meisenheimer \ea.
(1989), who explain their radio-to-optical spectra and infer the physical
conditions of the radiating plasma. Using synchrotron radiating electrons as a
trace, we can calculate the spectrum and the maximum energy of protons
accelerated under the same conditions. For simplicity, we disregard heavy
nuclei, but their probable role is discussed.  The normalization of proton
flux injected in extragalactic space is performed by using estimates from
Rawlings \& Saunders (1991) for the total energy stored in relativistic
particles inside the jets and radio galaxy evolution models given by Peacock
(1985).  We calculate the spectral modifications due to interactions of the
protons with the microwave background photons in an evolving universe,
following Berezinsky \& Grigor'eva (1988). Constraints on the extragalactic
magnetic field can be imposed, since it must permit an almost homogeneous
filling of the universe with energetic protons.  The observed ultra-high
energy cosmic ray spectrum is reproduced in slope and flux, limited at high
energies by the Greisen-cutoff at about $80\,$EeV. The requirements on the
content of relativistic protons in jets and the constraints to the
extragalactic magnetic field are consistent with common estimates.  The data
beyond the Greisen cutoff for protons may be explained by including heavy
nuclei in our model, since they can propagate over cosmological distances up
to more than $100\,$EeV.
\end{abstract}

\section{Introduction}

The contribution of extragalactic sources to the ultra-high energy part of the
cosmic ray spectrum (beyond the ``knee'' at about $5\,$PeV\footnotemark) is
still a matter of debate.  While it is easy to show that extragalactic sources
cannot provide enough energy to fill the whole universe with the total cosmic
ray (CR) energy density observed on earth, a contribution to the ultra-high
energy (UHE) part is possible, if the particle injection spectrum is
considerably flatter than the observed one.  This assumption implies no
contradiction, since at ultra-high energies, where the extragalactic cosmic
rays are believed to dominate over the galactic component, interactions of the
particles with the universal microwave background must be taken into account
(Greisen 1966, Stecker 1968, Blumenthal 1970), which in conjunction with the
cosmological evolution of the source density leads to a steepening of the
spectrum, as pointed out in detail by Hill \& Schramm (1985) and by Berezinsky
\& Grigor'eva (1988).  Additionally, a flat injection spectrum agrees with the
predictions of the theory of first order Fermi particle acceleration at
strong, non- or weakly relativistic shock waves (e.g.~Drury 1983).  Since the
hot spots in Fanaroff-Riley class \2 (FR-\2) radio galaxies are believed to be
the largest and most powerful shock waves in the universe, it is natural to
consider them as sources of high energy cosmic rays (Biermann \& Strittmatter
1987, Ip \& Axford 1991).

\footnotetext{$\rm1\,PeV\equiv10^{15}\,eV\,,\;1\,EeV\equiv10^{18}\,eV$.
Generally, we express astrophysical quantities in units commonly used for
them.}

Any model of the origin of the highest energy comic rays has to fulfill the
following conditions: (i) the predicted sources must be able to accelerate
particles to the observed energies up to at least $100\,$EeV, (ii) the
energy content in relativistic particles and the number density of the sources
must be sufficient to provide the observed UHE-CR flux, and (iii) the
observed UHE-CR spectrum must be fitted in the energy region where the
considered contribution dominates over all others. With respect to the hot
spots
in \FRII galaxies only point (i) has been discussed in the literature yet;
the aim of this work is to present a complete model which takes into account
all three conditions. The investigation of \FRII radio galaxies constitutes a
fairly good basis for this purpose: \FRII galaxies are very bright radio
objects, their distribution and evolution is well known up to rather high
redshifts (Peacock 1985), and their content of relativistic particles can be
well estimated from their sychrotron emission (Rawlings \& Saunders 1991).
The only problem is to obtain a reliable estimate of the proton to electron
ratio in the energy density of relativistic particles; this uncertainty
introduces as a ``fudge factor'' in our model, that can only be tested on
plausibility but not calculated at present.

Clearly, the hot spots in \FRII galaxies are not expected to be the only
extragalactic objects that can accelerate particles to ultra-high energies.
Most active galactic nuclei might provide much better conditions for that, and
shock acceleration can also take place in the jets of less luminous FR-I
galaxies (e.g. M87). The reason for the restriction to the rare class of FR-II
galaxies is, that highly energetic charged particles produced deep inside
galactic structures will suffer substantial adiabatic losses on their way out
to the extragalactic medium. This problem applies to all possible cosmic ray
sources except the FR-II hot spots, since they are located at the edge of
the extended radio lobe of the galaxy and the particles can immediately enter
the extragalactic space.  However, recent investigations show that also
protons from galactic cores can escape with very high energies, since they can
undergo isospin flips in $pp$ or $\gamma p$ reactions, leaving the galaxy as
an UHE neutron. At present it is not clear whether the resulting CR spectrum
extends to the highest energies considered here (Protheroe \& Szabo 1992,
Mannheim 1993).

Our model is based on the present knowledge about diffusive shock
acceleration, radio galaxy properties, galaxy evolution and cosmic ray
propagation. There are still a lot of open questions in this fields, which we
must cover by simplified assumptions. One of the main restriction is that {\em
we only consider the proton component of cosmic ray spectrum}; however, we
expect them to dominate because the gaseous halos of early type galaxies as
well as active galactic nuclei have cosmic abundances to within a factor of 3.
With respect to shock acceleration, we only use the canonical theory, leaving
room for theoretical refinements in following papers.

The paper is divided in the following parts: in \sect2 the maximum energy
which can be attained by first order Fermi acceleration under hot spot
conditions is investigated, taking into account energy losses of the
particles by synchrotron radiation and interactions with the ambient photon
fields, and particle losses due to the finite extension of the region where
diffusion takes place.  \Sect3 introduces the radio luminosity function of
\FRII galaxies (Peacock 1985) and discusses the total energy content of these
sources stored in relativistic protons and its connection to the radio
luminosity of the source, based on results given by Rawlings \& Saunders (1991)
for the case of the minimum energy condition.  \Sect4 describes the transport
of the particles through the universe, considering energy losses by pair- and
pion-production in interactions with the MBR while including the cosmological
evolution of this background, based on simple Friedmann models with vanishing
cosmological constant.  In \sect5 we present our results on the contribution to
the UHE-CR spectrum and its dependence on some poorly known parameters, the
role of prominent nearby \FRII galaxies and a possible cosmic ray anisotropy
at highest energies connected to their positions in the sky.  Finally, the
assumptions and restrictions of our model are discussed and
possible extensions are suggested.

An early analytical model about radio galaxy hot spots as sources of UHE
cosmic rays was first presented by one of us (P.L.B.) at a conference in honor
of M.M.~Shapiro in October 1990 in Washington, D.C. (Biermann 1990).  It was
already shown there that the expected total CR flux could be sufficient to
explain the observed events above about $1\,$EeV, but spectral details were
only discussed in a cosmologically local approximation. Some of the results of
the actual model were already presented on conferences at Ringberg Castle,
Germany, 1991, and at the Bartol Research Institute in Newark, Delaware, 1991
(Rachen \& Biermann 1992a,b).

\section{Particle acceleration in radio galaxy hot spots}
\subsection{Acceleration mechanism and hot spot properties}

The hot spots in strong extended radio galaxies, denoted as \FRII galaxies
following the classification of Fanaroff \& Riley (1974), are identified as
the endpoints of powerful jets ejected by the active nuclei of the galaxies
deep into the extragalactic medium (see, e.g., Meisenheimer \& R\"oser 1989).
It has been shown in several publications that the radio-to-optical spectra
emitted by those hot spots can readily be explained as synchrotron radiation
from particles accelerated at a strong shock wave by the first order Fermi
mechanism (e.g.~Biermann \& Strittmatter 1987, Meisenheimer \ea.~1989).

In this mechanism, particles are accelerated by repeated scattering back and
forth across a shock front in an at least partially turbulent magnetic field
(e.g.~Drury 1983).  At every turn, the particle gains a specific amount of
momentum, and on the other hand it has a certain chance to escape downstream
from the shock front, both only dependent on the upstream flow speed and the
compression ratio at the shock front.  This kind of stochastic acceleration
leads to a power law particle spectrum $\ii n(E) \propto E^{-\alpha_{\rm p}}$
for energies far above injection. In the simplest (canonical) case of a
non-relativistic shock and a main magnetic field parallel to the shock normal,
the index $\alpha_p$ only depends on the compression ratio: in a
non-relativistic gas we get $\alpha_p = 2$, for a relativistic gas we have
$\alpha_p = 1.5$ (Drury 1983).  In contrast, for highly relativistic shock
waves the spectrum of accelerated particles may strongly depend on the shock
velocity (Kirk \& Schneider 1987); numerical simulations of extreme
relativistic shocks in a totally tangled magnetic field yield a power law
spectrum with $\alpha_p = 3$ (Ballard \& Heavens 1992). For weakly
relativistic shocks the spectra may be flattened due to an increasing
compression ratio compared to the non-relativistic case (Heavens 1989, Ellison
1991); this effect can be substantially stronger at oblique shocks (Kirk \&
Heavens 1989, Takahara \& Terasawa 1991). Another refinement that might be
considered is the effect of second order Fermi acceleration at the
magnetohydrodynamic waves in the diffusion region, which also leads to flatter
spectra than inferred from the canonical theory (Kr\"ulls 1992).

The canonical theory of diffusive shock acceleration does not include any high
energy cutoff in the particle spectrum, since energy losses of the accelerated
particles are disregarded and an infinite extension of the diffusion region is
assumed.  In reality, of course, the diffusion region is always finite and
naturally synchrotron losses of the energetic charged particles in the
surrounding magnetic field must be considered. Biermann and Strittmatter
(1987, BS87) proposed a model which allows an explanation of the cutoff in
spectra of hot spots and AGN, usually of the order of $10^{14}\,$Hz, by
diffusive shock acceleration including synchrotron losses, if the turbulence
spectrum of the magnetic plasma follows a Kolmogorov law. In this model, the
magnetic turbulence is excited by energetic protons accelerated at the shock
front, and the outer scale of turbulence corresponds to the gyration radius of
the most energetic protons. For the protons a high energy cutoff of the order
of $10^{21}\,$eV is found if their energy is mainly limited by synchrotron
losses rather than by the finite range of the diffusion region or by
interactions with ambient photons. In its basic form the model assumes a
parallel shock wave and the diffusion coefficient is the same on both sides of
the shock. A modification of these assumptions can cause a change of the
inferred cutoff frequency by about an order of magnitude (e.g.~Jokipii 1987),
which allows an explanation of extraordinary spectra like that of Pictor A
west (see below) without changing the basic model.

Obviously, the synchrotron emission of the relativistic particles is nothing
else but the radio-to-optical spectra observed from the hot spots. For a power
law particle spectrum the synchrotron emission spectrum is also described by a
power law with a spectral index $\alpha_s = (\alpha_p - 1)/2$.  In five out of
six observed hot spots, Meisenheimer \ea.~(1989, M89) found radio spectral
indices of $\alpha_s \simeq 0.5$ within the error estimates, which leads to
$\alpha_p \simeq 2.0$ in perfect agreement with the predictions of first order
Fermi acceleration at strong, non- or mildly relativistic parallel shock waves
in a nonrelativistic gas.  The only exception, Pic~A~west, shows
$\alpha_s\simeq 0.39$ and therefore requires flatter particle spectra
($\alpha_p\simeq 1.78$).  It should be stressed that all these hot spot
spectra are considerably flatter than the radio spectra of the extended
emission regions of the galaxies ($\exval{\alpha_s}\approx 0.8$).  In four
cases, their observations also confirmed the cutoff frequencies of about
$10^{14}\,$eV in perfect agreement with BS87, but also the remaining cases fit
into the prediction range of this model. Meisenheimer \ea.~used a variety of
prescriptions to estimate the magnetic field in hot spots, minimum pressure
was just one such method, the location of the loss bend in the synchrotron
emission spectrum another.  They found general agreement between these various
methods for mildly relativistic electron-proton jets and sub-milli-Gauss
magnetic field strengths in the hot spot region (see \tab{hotspots}).

In conclusion, the simple canonical theory of diffusive shock acceleration
explains the main hot spot features and is therefore a very attractive model
for the acceleration of cosmic ray particles.  Clearly, slightly flatter
spectra as predicted by enhanced acceleration models are consistent with most
of the results within the error estimates, even required by the observations
of Pic~A~west.  Thus, in our calculation we will investigate the influence of
a variation in the mean particle injection spectral index on our final
results.
\begin{footnotesize}
\begin{table}
\caption[]{\small\label{tab_hotspots} Hot spot properties,
taken from Meisenheimer \ea.~1989 (M89)}
\begin{flushleft}
\begin{tabular}{llllr@{\ }lll} \hline
Hot-Spot  & Radius$^*$     & Length$^*$     & Spectral index        &
Synchrotron & cutoff & Magnetic field$^*$
                   & Jet speed$^*$    \\
          & $R_{\rm HS}\;$[kpc]& $L_{\rm HS}\;$[kpc]& $\alpha_s$ (radio)    &
$\nu_c\;$[Hz]      & & $B_{\rm HS}\,$[mG]
                   & $\beta_{\rm jet}$ \\
\hline
3C20 West & $0.38\pm 0.02$ & $0.13\pm 0.08$ & $0.53\pm 0.05$ &
$1.6\cdot 10^{14}$ & & $\ii 0.45\vphantom{\Big|}^{\,+0.29}_{\,-0.12}$
                   & $0.51\pm 0.10$ \\
3C33 South& $0.66\pm 0.02$ & $0.06\pm 0.03$ & $0.59\pm 0.08$ &
$2.8\cdot 10^{14}$ & & $\ii 0.37\vphantom{\Big|}^{\,+0.16}_{\,-0.07}$
                   & $0.24\pm 0.04$ \\
3C111 East& $1.00\pm 0.02$ & $0.07\pm 0.07$ & $0.54\pm 0.05$ &
$8.5\cdot 10^{13}$ & & $\ii 0.20\vphantom{\Big|}^{\,+0.22}_{\,-0.07}$
                   & $0.35\pm 0.12$ \\
3C123 East& $\sim 2.5$     & $\sim 4$       & $0.52\pm 0.05$ &
$< 6\cdot 10^{13}$ & & $\ii 0.22\vphantom{\Big|}^{\,+0.06}_{\,-0.06}$
                   & $0.34\pm 0.05$ \\
Pic A West& $0.49\pm 0.02$ & $1.07\pm 0.05$ & $0.39\pm 0.02$ &
$> 7\cdot 10^{15}$ & & $\ii 0.36\vphantom{\Big|}^{\,+0.04}_{\,-0.04}$
                   & $0.16\pm 0.03$ \\
3C273 A & $1.15\pm 0.4 $ & $2.0\pm 1.0$ & $0.45\pm 0.10$ &
$1.0\cdot 10^{14}$ & & $\ii 0.54\vphantom{\Big|}^{\,+0.20}_{\,-0.10}$
                   & $0.21\pm 0.04$ \\
\hline
\end{tabular}\\[0.5ex]
$^*\;$Valid for $q_0=0$ and $h_0=0.5$ ($H_0 = 100\,h_0\,\rm
km\,sec^{-1}\,Mpc^{-1}$). See M89 how the parameters depend on cosmology.
\end{flushleft}
\end{table}
\end{footnotesize}

\subsection{Energy limits for proton acceleration}
\subsubsection{Energy losses}

The simple power law spectra discussed above are steady state solutions and do
not include the usually energy dependent time scale $\tau_a$ for the
particles to get accelerated. This time scale is strongly dependent on the
turbulent energy spectrum of the magnetic field, $I(k)$, i.e.~the magnetic
energy density per unit wavenumber $k$. Following BS87, we assume a
Kolmogorov spectrum $\ii I(k)\propto k^{-5/3}$ and get in the canonical case
	\beq
	\label{tau_+}
	\tau_+(\Ep)\,=\,\frac{20 \kappa_\|}{c^2\beta_{\rm jet}^2}\quad,\qquad
		\kappa_\| \,=\, \frac{2c\,R_{\rm T}^{\;2/3}}{\pi b}\,
            			\left(\frac{\Ep}{eB}\right)^{\TS\frac{1}{3}}
	\quad,\eeq
with a total magnetic field strength $B$, a fraction $b$ of the magnetic
energy density given in the turbulent field integrated over all wavenumbers
and an outer scale of turbulence $R_T$. We have to use effective quantities in
\eq{tau_+}, averaged about upstream (jet) and downstream (hot spot)
conditions, noting that downstream counts a fraction of $4/5$ (see Drury
1983).

On the other hand, energetic protons lose energy by synchrotron radiation
and by interactions with the ambient photon fields over a time scale
$\tau_-$, given by
	\beq\rn
	\label{tau_-}
	\tau_-(\gp) \,=\, \frac{6\pi\,m_p^3 c}{\sigma_T m_e^2 B^2 (1+Aa)}\,
	\frac{1}{\gp}\quad,
	\eeq
where $\sigma_T$ and $m_e$ are the Thomson cross section and electron mass,
respectively, $\ii a = U_\gamma/U_B$ is the energy density of the ambient
photon field relative to that of the magnetic field, and $A$ is a quantity for
the relative strength of $p\gamma$ interactions compared to sychrotron
radiation (see BS87). In hot spots we have typically $A \approx 200$ and
$a\gsim 0.01$; an upper limit for $a$ cannot be given since the photon field
in the hot spot may be highly anisotropic due to a boosted radiation cone from
the core.

Obviously, acceleration can only work as long as $\tau_+ < \tau_-$, so the
cutoff energy $E_c$ is given by $\tau_+(E_c) = \tau_-(E_c)$.  For typical hot
spot conditions, i.e. $B \approx 0.5\,$mG and $\beta_{jet} \approx 0.3$, and
assuming the outer scale of turbulence to be connected to the hot spot radius
($R_T\approx 1\,$kpc), we get $E_c > 100\,$EeV for $a < 0.1$. In the self
consistent model suggested by BS87, where $R_T$ is given by the gyration
radius of the highest energy particles, $E_c$ approaches $10^{21}\,$eV. Hence,
the consideration of energy losses alone allows protons to be accelerated to
the highest observed CR energies in a typical \FRII hot spot.

\subsubsection{Diffusive particle losses}

The canonical theory of diffusive shock acceleration does not only assume an
infinite time, but also an infinite space for the particles to diffuse in the
medium around the shock front; the particles can only escape due to their bulk
movement with the plasma flow in downstream direction.  A limitation of this
space will lead to an additional, energy dependent loss of particles by
diffusion over the outer boundaries of the system.  The strongest limit is
given by the requirement that the gyration radius $r_g = \Ep/eB$ of the
particle must be smaller than the extension of the magnetic environment.
Assuming that the magnetic field of the hot spot is limited to the observable
region and taking the physical quantities as above, we get an energy limit
$\Ep < 500\,$EeV by that condition.  However, the shock structure in a hot
spots is likely to be much more extended than visible in the nonthermal
radioemission, as suggested by magnetohydrodynamical modelling, involving a
central Mach disk, and biconical oblique shocks around it; hence we suspect
that the sizes of the emission regions which are tabulated in \tab{hotspots},
are actually lower limits to the physical size of the region involved in the
acceleration of protons. On the other hand, diffusive particle losses may
modify the spectrum also at lower energies.

To estimate the critical energy above which diffusion particle losses will
alter the spectrum, we can argue similarly as in the case of energy losses.
First, we distinguish between lateral and parallel diffusion, respective to
the main direction of the magnetic field, which we assume to be mostly
parallel to the jet axis. Lateral diffusion is only important for diffusive
particle losses, while parallel diffusion governs the acceleration process.
To lowest order, the timescale $\tau_\perp$ for lateral diffusive particle
losses is approximately given by
	\beq\rr
	\label{tau_perp}
	\tau_\perp \approx \frac{R_{HS}^2}{5\kappa_\perp}\quad,
	\eeq
with $\kappa_\perp$ being the diffusion coefficient perpendicular to the mean
magnetic field (e.g. Forman \& Webb 1985). In the quasilinear regime, we
always have $\tau_+/\tau_\perp \propto \kappa_\|\kappa_\perp = (\frac{1}{3}r_g
c)^2$, so that $\tau_+ = \tau_\perp$ leads to a critical energy $E_{c\perp}
\approx 40\,$EeV using typical hot spot conditions as above, which is totally
independent of the magnetic turbulence spectrum.

An additional limit turns up through the length of the hot spot, in
other words, the extension of the downstream region parallel to the mean
magnetic field.  The parallel diffusion, which takes the particles repeatedly
across the shock front, may also lead them across the outer border so that the
particles get lost.  In the transport equation, the situation can be treated
by taking $\kappa_\|$ space dependent, falling rapidly to zero at some
distance from the shock front.  However, for our estimate of the critical
energy it is sufficient to compare the the hot spot length $L_{HS}$ with
diffusion length given by
	\beq
	\label{difflen}
	l_\| \approx \frac{4\kappa_\|}{c\beta_{\rm jet}}\quad,
	\eeq
for strong shocks. $l_\|$ gives the distance from the shock front within most
of the particles are scattered back to the shock, and the turbulence spectrum
becomes important again; taking a Kolmogorov spectrum, we are led to a
critical energy $E_{c\|} \approx 1\,$EeV for a hot spot length $L_{HS}\approx
R_T \approx 1\,$kpc. Assuming $R_T\approx 0.2\,$kpc, corresponding to the
gyration radius of $100\,$EeV protons, yields $E_{c\|} \approx 30\,$EeV.

Hence, diffusive particle losses seem to have the strongest influence on the
energy distribution of protons accelerated by diffusive shock acceleration in
hot spots, as already argued by Ip \& Axford (1991).  For parallel losses we
find a critical energy of 1\,EeV, but we rather expect a slight modification
of the spectrum near to that energy than an abrupt cutoff.  For instance, in
case of a Kolmogorov turbulence spectrum the energy dependence of the
diffusion length is weak ($\ii l_\| \propto E^{1/3}$). Lateral diffusion losses
can give a cutoff, but the critical energy here is near 40\,EeV. Obviously,
both estimates depend strongly on the acceleration time scale, which can be
smaller up to an order of magnitude if we consider oblique shocks, thus the
cutoff energy could be shifted by the same amount (Jokipii 1987, Takahara \&
Terasawa 1991).  Moreover, the plasma may not be adequately described as
quasilinear; the magnetic field structure behind supernova shocks is one well
known example, where the radio polarization shows unambiguously (Dickel
\ea.~1991) that the magnetic field structure is radial in the shock region.
Here convective motions appear to dominate the turbulence, and thus
the corresponding diffusion is likely to be very different from a
quasilinear approximation (see Jokipii 1987, Biermann 1993).

The only limit independent of the shock obliqueness and the plasma properties
is given by the gyration radii of the particles, but even for the lowest
estimated magnetic field strength in hot spots this is above 100\,EeV. Thus it
is reasonable to assume that the acceleration of protons to highest CR
energies in hot spots is possible. In the following discussion we will use
$10^{21}\,$eV as a strict upper limit for the particle injection energy, but of
course we will investigate the influence of a lower injection cutoff between
30 and 300\,EeV on our final results.

\section{Source function of cosmic rays}

\subsection{Radio galaxy classification}

\FRII galaxies are extended, steep spectrum radio galaxies, that are
uniquely classified by their double structure due to their luminous hot spots,
which makes them interesting as cosmic ray sources. On the other hand, this
morpology occurs for radio luminosities above $2\cdot 10^{26}\,
\rm W Hz^{-1}$
at
178 MHz, which makes it easy to distinguish \FRII galaxies from other radio
galaxies in the radio luminosity function (see 3.2).  The present state of
knowledge is that roughly $70\%$ of all steep spectrum radio sources in this
power range show \FRII structure (Perley 1989), but on the other hand there
may also be some compact flat spectrum sources that can be identified with
\FRII galaxies having their jets oriented along the line of sight (Padovani \&
Urry 1992). At high redshifts (above z=1), however, little is known about the
morphology of the sources and it is not clear whether the Fanaroff-Riley
relation still holds; but we will see in \sect4 that this uncertainty affects
the cosmic ray spectrum only below 1\,EeV.

\subsection{Kinetic jet energy and radio luminosity}

The synchrotron radiation emitted by relativistic electrons is the only (or at
least the main) information we get about the content of relativistic
particles in jets. However, the total sychrotron luminosity $L_s$ of a source
allows us to give minimum estimates about their mean energy density in
relativistic particles $U_p$ and $U_e$, and in the magnetic field $U_B$. The
minimum energy condition is given by (see, e.g., Pacholczyk 1970)
	\beq\rr
	\label{U_min}
	U_B = \TS\frac{3}{4}\,(1+k_p)\,U_e\quad.
	\eeq
Here we introduce the quantity $k_p = U_p/U_e$, where $U_p$ is the energy
density of {\em relativistic} protons in the source. The role of the protons
is double-edged here: They do not emit any synchrotron radiation in
the frequency band usually considered, but they influence the energy balance
at minimum. Since we have no information about them, $k_p$ must be considered
as an unknown quantity, for which only plausibility limits can be given (see
below). However, the emission of photons in the X and $\gamma$ regime from UHE
protons in hot spots by synchrotron radiation and the proton-induced
electromagnetic cascades can by expected (Mannheim \& Biermann 1989, Mannheim
\ea.~1991). Recent GRO detections of blazars, which may be considered as
compact Doppler-boosted emission from jets support this expectation (Mannheim
\&
Biermann 1992, Mannheim 1993) and give evidence for the presence of
energetic protons employed in this work. In the future, if observational
material on X-ray and hard $\gamma$ emission from hot spots can be obtained,
it may even be possible to derive definite estimates on $k_p$.

It seems important to define the term ``relativistic'' in the current context:
It depends on the frequency band of the observed sychrotron radiation; for
example, a lower frequency limit of about $100\,$MHz corresponds to $\gamma_e
\ga 300$.  However, the total synchrotron luminosity is usually determined by
extrapolating the observed spectrum down to zero, so we are nearly consistent
if we extend the inferred particle spectrum down to $\gamma = 1$ (neglecting
nonrelativistic corrections). Clearly, the error in normalization
introduced by this approximation is larger for the steeper particle spectrum
than for the synchrotron spectrum, but in principle any additional population
of nonrelativistic particles in the jet (i.e.~the plasma flow) has no effect
on our energy balance, and the estimates used here give no limit to the {\em
total}\/ kinetic energy of the jet.

\subsection{Proton content of the jets}

For considering \FRII galaxies as CR sources, we are only interested in the
``proton luminosity'' of a source. From the synchrotron luminosity one can
derive the total jet power, assuming some value for $k_p$. Rawlings \& Saunders
(1991, RS91) gave the jet power for $39$ \FRII galaxies, assuming no protons in
the jet ($k_p = 0$); in the following, we will denote this jet power for the
no-proton-case as $P^{(0)}_{jet}$. From the minimum energy condition it follows
for any non vanishing $k_p$
	\beq\rn
	\label{Pjet}
	P_{jet}(k_p) = (1+k_p)^{\mbox{$\ii\SS 4/7$}}\,P_{jet}^{(0)}\quad.
	\eeq
With help of \eq{U_min} we easily get
	\beq\rn
	\label{fudge}
	L_p \equiv f(k_p)\,P_{jet}^{(0)}\quad,\qquad
	f(k_p) = \TS {4\over 7}\,k_p\,(1+k_p)^{\mbox{$\ii\SS -3/7$}}
	\quad,\eeq
calling $f$ the {\em fudge factor} in our normalization. Later, we will derive
$f$ from fitting our calculated spectra to the data and therefore the relative
proton content $k_p$ required.

To test our requirements on plausibility we can give an upper limit on $k_p$
by considering the total power supplied by the central engine: RS91 derived
minimal kinetic jet powers up to $10^{39}\,$W assuming no relativistic
protons, but the main part of the total energy may lie in the thermal flux
rather than in the relativistic particles. From present radio data it is
suggested that the total power supply from active galactic nuclei does not
exceed $10^{41}\,$W, thus we should consider $k_{p(max)}\approx 1000$ as an
upper limit by order of magnitude.  A lower limit estimate on $k_p$ can be
derived by assuming an equal normalization of both the electron and proton
spectra at the injection: here, we simply get $k_p = \ln\gamma_{c(p)}\, /
\,\ln\gamma_{c(e)}$ for an $E^{-2}$ spectrum. For $\gamma_{c(p)}\approx
10^{11}$ and $\gamma_{c(e)}\approx 10^5$ this yields $k_p\approx 2$, as
already used by M89.  An independent estimate is given by comparison to the CR
spectrum observed in the Galaxy, where $k_p\approx 100$ is measured for
particle energies above $1\,$GeV; extrapolating to lower energies implies
smaller values. Now, galactic cosmic rays can also be assumed to be
accelerated at shock waves, but clearly the environment is quite different, so
it is not clear how far we can apply this galactic quantity to hot spots. In
the following, we will use the regime between $2$ and $100$ as the confidence
interval for $k_p$, although we stress that these limits are not strict.

\subsection{Application of radio luminosity functions}

\begin{figure}[tp]
\vspace{9.5cm}
\caption[]{\small\label{fig_jet:radio} Correlation between specific radio
luminosity and jet power for \FRII galaxies; data from RS91 and regression
line (\ref{jet:radio})}
\end{figure}
To calculate the extragalactic source function of CR protons, we have
to apply our knowledge about jet powers and proton luminosities to the epoch
dependent radio luminosity function (RLF) of \FRII galaxies.  The RLF $\ii
n_{\rm G}(P_\nu,z)$ gives the number density of radio galaxies with a
specific radio luminosity $\ii P_\nu$ at a given frequency $\nu$ at an epoch
$z$ respective to the {\em comoving} unit volume element. RLF's can be
derived from radio source counts, using the available information about
redshifts and assuming some modelling parameters (Peacock \& Gull 1981,
Windhorst 1984, Peacock 1985).  Clearly, RLF's derived this way depend on the
choice of cosmology, but it is quite simple to change to another cosmological
model
by using the relation
	\beq
	\label{RLF_cosvar}
	n_{\rm G(1)}(P_{\nu(1)},z)\,\frac{dV_{\rm c(1)}}{dz} \,=\,
	n_{\rm G(2)}(P_{\nu(2)},z)\,\frac{dV_{\rm c(2)}}{dz} \quad,
	\eeq
noting that $\ii P_\nu\,d_{\rm L}^{-2}$ is an observational quantity (the
brightness of the source) and does therefore not depend on cosmology (Peacock
1985).

To connect the proton source function to the RLF, the given relation between
{\em total\/} synchrotron luminosity and kinetic jet energy is not very
useful; we rather have to find a relation between jet energy and radio
luminosity at the frequency for which the RLF is given. In this work, we use
the luminosity functions RLF~1 to RLF~4, given by Peacock (1985) for various
modelling parameters, based on source counts at $2.7\,$GHz for steep spectrum
radio sources.  Hence, we compare $P^{(0)}_{jet}$ to $\ii\hat P_{2.7}$, the
radio luminosity per unit solid angle at $2.7\,$GHz, for the 39 sources
investigated by RS91; \fig{jet:radio} shows a very good correlation, the
correlation coefficient derived from linear regression is $0.8$.  For the
regression line we find
	\beq \label{jet:radio}
	\log\big(P_{\rm jet}^{(0)}\,\big/\,{\rm Watt}\big) =
		15.5 + 0.63\,\log\big([\nu P_\nu]_{2.7}\,\big/\,\rm Watt\big)
	\quad.\eeq
Note that \eq{jet:radio} does not depend on cosmology, since it expresses a
relation between total luminosities.  Multiplying with our fudge factor
$f(k_p)$, we directly have the averaged proton emission for a \FRII galaxy as
a function of $\ii\hat P_{2.7}$.  The proton source injection function
$\Psi(z)$ is now given by
	\ii\beqar
	\label{psi}
	\Psi(E,z) &=& f(k_{\rm p})\,A(L_{\rm p})\,E^{-\alpha_{\rm p}}\,C(E) \\
	  &\times& \int\limits_{\hat P_{2.7\rm (min)}}^{\hat P_{2.7\rm (max)}}
	      	   \!\!\!\d \hat P_{2.7} \;P^{(0)}_{\rm jet}(\hat P_{2.7})\;
			 n_{\rm G}(\hat P_{2.7},z)\quad.\nonumber
	\eeqar\ri
Here we introduced the cutoff-function $C(E)$ to describe the modification of
the pure power law spectrum due to cutoff effects; we will not try to give
$C(E)$ a concrete form in this paper, instead we will use an exponential
cutoff as a working hypothesis in \Sect5.  The lower limit of integration is
given as $\ii\hat P_{2.7\rm(min)}\approx 2\cdot
10^{24}\rm\,W\,Hz^{-1}\,ster^{-1}$ by our restriction to consider only \FRII
galaxies. The upper limit is given as $10^{30}\,\rm W\,Hz^{-1}\,ster^{-1}$ by
the validity of Peacock's model fits. The uncertainties in the Fanaroff-Riley
relation between morphology and radio luminosity are small compared to the
fudge factor $f(k_p)$.  The factor $A(L_p)$ is found by normalizing the proton
spectrum to the total luminosity $L_p$, where it is sufficient to use the
unmodified power law spectrum with an abrupt cutoff ($A = L_p/\ln\gamma_c$ for
$\alpha_p = 2$). \Fig{evol} shows the normalized, epoch dependent proton
injection $f^{-1}\int \Psi(E,z)\d E$ for the various RLFs, making clear that
there are no big uncertainties in it up to redshifts near $z=2$, if we assume
that the Fanaroff Riley relation still holds at high redshifts.
\begin{figure}[tp]
\vspace{9.5cm}
\caption[]{\small
\label{fig_evol} Evolution of the total cosmic ray power injected
by \FRII galaxies. Straight lines correspond to an evolution $\propto (1+z)^m$
for integer values of $m$ ($h_0=0.75$, $q_0=1/2$). The radio luminosity
functions RLF~1-4 used here are given by Peacock (1985)}
\end{figure}

\section[]{Cosmological modification of the cosmic ray \\ spectrum}

Protons with Lorentz factors $\gp \gsim 10^{10}$ can undergo high energy
reactions even with the very low energy photons of the universal microwave
background radiation (MBR), as pointed out first by Greisen (1966) and
independently by Zatsepin \& Kuzmin (1966). The attenuation length reduces to
merely $10\,$Mpc for $\gp \gsim 10^{11}$ (Stecker 1968), what causes a cutoff
in the particle spectra of distant extragalactic sources, usually denoted as
the {\em Greisen-cutoff}\/. Hillas (1968) demonstrated the influence of the
cosmological evolution on this effect; for very distant sources the
Greisen-cutoff apears at much lower energy, because of the higher density and
temperature of the MBR at earlier epochs.  The most extensive investigations
of the transport of UHE protons in the evolving MBR were done by Hill \&
Schramm (1985, HS85) and by Berezinsky \& Grigor'eva (1988, BG88), who both
considered the modification of the spectrum due to the cosmological evolution
of the proton sources.  While HS85 expended a lot of work on integrating the
exact transport equation numerically, treating the various particle reactions
in very great detail, BG88 presented a very instructive way of simplifying the
transport equation, which allows a proper treatment of various cosmological
models. The results of both can be found to be in fair agreement, if the
different treatment of cosmology and the different boundary conditions in the
proton injection are properly considered (as discussed below). A purely
analytical method was suggested by Stecker (1989), which however only allows a
cosmologically local treatment of the problem. We will use the method of
Berezinsky and Grigor'eva in the following.

\subsection{Proton--photon interactions}

Since the proton is the lightest baryon it cannot be destroyed by particle
reactions; it will always reveal itself after a number of decays of heavier
particles, that are produced in the reaction.  However, particle reactions
will always cause a more or less substantial energy loss of the proton.  The
crucial quantity expressing the importance of a $\gamma p$ reaction in
modifying proton spectra is the product of its cross section $\sigma$ and its
inelasticity $\kappa = 1 - E_f/E_i$, where $E_i$ and $E_f$ are the protons
initial and final energy, respectively (Stecker 1968).  It turns out, that
only photomeson production and $e^+e^-$ pair production in interactions of UHE
protons with the MBR modify the spectrum considerably.
Compton scattering can be neglected as well as the interaction with other
photon backgrounds.  Only the infrared background caused by dust emission of
early galaxies may have an influence at early epochs (Puget \ea.~1975), which,
however, will not be so large that it could change our results noticeably.

\subsubsection{Photomeson production}

Photomeson production, i.e.~the production of one or more pions or other
mesons by $\gamma p$ reactions, is the dominant process for limiting the
travel time of UHE protons in extragalactic space, since its inelasticity is
rather high.  It sets in at a photon energy in the protons rest frame (PRF) of
$145\,$MeV with a strongly increasing cross section at the $\Delta(1232)$
resonance, which decays into the one-pion channels $\pi^+n$ and $\pi^0p$.
Here, we do not distinguish between neutrons and protons, since the particles'
charge is not important in this context and the inelasticity introduced by the
beta decay of the neutron is negligible.  At higher energies, heavier baryon
resonances occur and the proton might reappear only after successive decays of
resonances; the most important channel of this kind is $p\gamma\to\Delta^{++}
\pi^-,\;\Delta^{++}\to p\pi^+$, which is the main decay channel of the
$\Delta(1620)$ and $\Delta(1700)$ resonances.  With increasing energy the
reactions are no more mediated by resonances, as the diffractive production of
the vectormesons $\rho$ and $\omega$ or direct multi-pion production.

The inelasticity of a reaction depends not only on the outgoing particles but
also on the kinematics of the final state. However, for most reactions we can
find a reasonable way of averaging about final state kinematics. For
reactions mediated by resonances we can assume a decay, which in the center
of mass frame is symmetric in forward and backward directions with respect to
the reaction axes (given by the incoming particles). In this case, the mean
inelasticity of a reaction, that reveals the proton after $n$ resonance
decays is given by
	\beq
	\label{inel1}
	\exval{\kappa} \,=\, 1 - {1\over 2^n}\,\prod_{i=1}^n
	\hbox{$\DS\ri\left(1 + \frac{m_{R_i} - m_{M_i}}{m_{R_{i-1}}}\right)$}
	\quad,\eeq
where $m_{R_i}$ denotes the mass of the $i$-th resonance in the decay chain
and $m_{M_i}$ the mass of the associated meson; by definition, we set
$m_{R_0}c^2 \equiv \sqrt{s}$, the total energy of the reaction in the center
of mass frame (CMF), and $\ii R_n$ is actually the proton again; for all other
resonances, we use their nominal masses.  For single-pion production,
\eq{inel1} reduces to the formula originally given by Stecker (1968).  On the
other hand, diffractive reactions such as $p\gamma\to p\rho^0$ can be treated
as strictly forward peaked, i.e.  $\chi = 0$ for the scattering angle in the
CMF.  Denoting the particle speed in the CMF with $\beta^*$, this adds a
factor $(1+\beta^*_{R_1})$ to the product in \eq{inel1}; therefore, at
threshold ($\beta^* = 0$) the inelasticity of diffractive scattering is the
same as for resonance reactions, but it decreases strongly with increasing $s$
($\beta^*\to 1$).  If there are no intermediate resonances, i.e. for the
reactions $p\gamma\to p\pi^0$ (resonance) and $p\gamma\to p\rho^0$
(diffractive), for $s\to\infty$ we get $\exval{\kappa}\to 0.5$ in the
resonance case and $\exval{\kappa}\to 0$ in the diffractive case.

For multi-pion production the case is much more complicated because of the
nontrivial final state kinematics.  However, it is known by experiment, that
for very high $s$ the incoming particles lose only half of their energy for
the production of pions, no matter how many pions are produced -- this is
known as the {\em leading particle effect}.  On the other hand, the
inelasticity of $n$-pion production {\em at threshold} is given by
\eq{inel1}; if we assume that at all energies multi-pion channels at
threshold always dominate about channels far from threshold (behavior as
diffractive scattering), it appears reasonable to use the following
approximation for the total multi-pion inelasticity:
	\beq
	\label{inel2}
	\exval{\kappa}_{\rm multi} \;\approx\; \kappa_{\rm th}(\exval{n_\pi})
	\;=\;\frac{\exval{n_\pi} m_\pi}{m_{\rm p} + \exval{n_\pi} m_\pi}
	\quad.\eeq
Here $\ii\exval{n_\pi}$ is the {\em net} mean pion multiplicity; this is the
mean multiplicity over all channels, which can be taken as a function of $s$
from the analysis of cosmic ray events in the atmosphere or from accelerator
data (Particle Data Group 1988), reduced by the contribution of channels we
treat separately. \Eq{inel2} may be a rather crude simplification, but it
fulfills the boundary conditions given by the leading particle effect for
energies up to $200\,$GeV photon PRF energy, and gives the correct value for
the inelasticity at the threshold for two-pion production. In the
astrophysical scenario multi-pion production is of minor importance anyway,
thus more effort in calculating inelasticities at this point would be
unreasonable. \Fig{inel1} shows the quantity $\exval{\sigma\kappa}$ for
several reaction channels, where all channels not treated separately are
sampled in the multi-pion channel.  Obviously, photomeson production sets in
very rapidly and leads to an energy loss constant by order of magnitude up to
highest energies.
\begin{figure}[t]
\vspace{9.5cm}
\caption[]{\small\label{fig_inel1} Cross section weighted by inelasticity for
different channels of photomeson-production ($M\pi$ is the multi-pion channel,
see text)}
\end{figure}

\subsubsection{Pair production}

Even at energies two orders lower than required for photomeson production,
electron-positron pairs can be created by MBR photons at UHE protons.  In
the PRF, the photon dissociates into a lepton pair, transferring only a small
amount of momentum to the proton; in the first Born approximation, the energy
transfer is treated as zero.  However, in the lab frame the energy necessary
for the production of the $e^+e^-$ pair is supplied nearly alone by the
proton, leading to an inelasticity roughly equal to the mass ratio
between the pair and the proton.
\begin{figure}[tb]
\vspace{9.5cm}
\caption[]{\small
\label{fig_inel2} Cross section weighted by inelasticity for
pair

production}
\end{figure}

In contrast to photomeson production, the exact energy losses caused by pair
production can be calculated purely theoretically, since this reaction is
governed by quantum electrodynamics.  We calculate $\exval{\sigma\kappa}$ for
pair production in the first Born approximation, following Blumenthal (1970);
the result is shown in \Fig{inel2}.  While the cross section of pair
production is more than an order of magnitude higher than that of photomeson
production, $\exval{\sigma\kappa}$ is lower because of the much lower
inelasticity; nevertheless, pair production is of extraordinary importance for
the transport of UHE protons, since it sets in at lower energy and can
therefore influence protons that cannot undergo pion production.  This effect
is important in thermal photon fields like the MBR and should not be confused
with the situation in hard photon spectra, where for all proton energies there
are enough photons available to produce pions, which is of course much more
effective.

\subsection{Energy evolution of cosmic ray protons}

\begin{figure}[t]
\vspace{9.5cm}
\caption[]{\small\label{fig_Ez} Energy evolution of cosmic ray protons
($h_0=0.75$, $q_0=1/2$)}
\end{figure}
Clearly, interactions of UHE protons with MBR photons cause a step-by-step
energy loss, which should be considered as a collision integral in the
transport equation. However, if every single step causes only a small energy
loss and a lot of interactions take place on the way of the proton through
the universe, we can approximately treat the evolution of the proton's energy
as continuous. This approximation is quite good for pair production, and
leads to qualitatively correct results for photomeson production, as shown by
BG88.  Hence, we introduce a function $\eta(z) = -1/E\;(\d E(z)/\d t)$,
where $z$ denotes the cosmological epoch in terms of its corresponding
redshift. Now, the evolution function of the proton energy $E(z)$ is given as
the solution of an ordinary differential equation
	\beq
	\label{Ep_DGL}
	\frac{1}{\Ep}\,\frac{d\Ep}{dz} \;=\; \frac{1}{1+z} +
	\frac{1+z}{H_0\,\sqrt{1 + 2q_0 z}}\;\eta_{\rm 0(mbr)}(\gp [1+z])
	\quad.\eeq
Here we adopt Friedmann cosmology with vanishing cosmological constant
$\Lambda$, $H_0$ and $q_0$ denoting Hubble's constant and the deceleration
parameter at present epoch, respectively.  Though both parameters are poorly
known we will see in \sect5 that only a variation of $H_0$ gives a
considerable effect on the resulting cosmic ray spectrum; therefore, we expect
that a extension to cosmological models with $\Lambda\neq 0$ would not produce
remarkable changes.  $\eta_{0(mbr)}(E_p)$ denotes the energy loss rate caused
by interactions with the MBR at present epoch ($z = 0$), which is given by
(BG88)
	\ii\beqar
	\label{eta_mbr0}
	\eta_{\rm 0(mbr)}(\gp) &=&
	\frac{-k T_0}{2\pi^2\hbar^3 c^2}\,\frac{1}{\gp^2} \\
	&\times& \sum_j\!\int\limits_{ \epsilon'_{j,\rm th}}^{\infty}\!\!
	\d\epsilon'\,\epsilon'\exval{\sigma\kappa}_j(\epsilon')\;
	\ln\left[1 - \exp\left(\frac{-\epsilon'}{2\gp k T_0}\right)\right]
	\nonumber\eeqar\ri
Here, the sum is carried out over all reaction channels, integrating over the
photon PRF energies $\epsilon'$ up from threshold.  $T_0$ is the present
temperature of the MBR, recently measured as $2.735\,$K by the COBE and COBRA
experiments in good correspondence (Gush \ea.~1990, Mather \ea.~1990).  The
dependence of $\eta_{mbr}$ on $z$, caused by the evolution of the MBR, is
	\beq\rr
	\label{eta_mbr(z)}
	\eta_{mbr}(\gp,z) \,=\, (1 + z)^3\,\eta_{0(mbr)}
	\left(\gp [1+z]\right)\quad,
	\eeq
The first term on the left hand side of \eq{Ep_DGL} derives from the adiabatic
energy loss of the protons itself in an expanding universe, which can be
written as
	\beq\rr
	\label{eta_ad(z)}
	\eta_{ad}(z) \,=\, (1 + z) \sqrt{1+2q_0 z}\;\,H_0
	\quad.\eeq
Obviously, $\eta_{ad}(z)$ is nothing else than the redshift dependent Hubble
parameter.

\Fig{Ez} shows the evolution of proton energy with $z$, governed by
\eq{Ep_DGL}, for various values of its present (observed) energy.  The figure
shows clearly the turnover from the linear, adiabatic decrease of energy to an
exponential decrease caused by pair production, for protons of {\em any}
observed energy $E_0$ at some epoch $z_e$.  Going further back in time,
photomeson production sets in at an epoch $\ii z_\pi$ and causes an energy loss
so rapid that unreasonable high injection energies would be required to produce
a proton observed with an energy $E_0$ at any epoch significant earlier than
$\ii z_\pi$.  \Fig{zep} shows $z_e$ and $\ii z_\pi$ as a function of $E_0$, the
latter can be considered as the border of the {\em confinement volume}
transparent to CR protons observed at $E_0$.
\begin{figure}[p]
\vspace{9.1cm}
\caption[]{\small\label{fig_zep} Dominant energy loss processes for cosmic ray
protons observed with energy $E_0$ during cosmological evolution}
\end{figure}
\begin{figure}[p]
\vspace{9.1cm}
\caption[]{\small
\label{fig_tau} Travel time for a proton started with
$10^{21}\,$eV

and observed with energy $E_0$}
\end{figure}

Finally, \fig{tau} shows the travel time of a proton injected with $E_i =
10^{21}\,$eV and observed with energy $E_0$, for various values of the
cosmological parameters $H_0$ and $q_0$. Clearly, only for low energy, where
adiabatic losses dominate, a dependence of cosmology introduced by the
redshift-time relation becomes noticable. However, from \eq{RLF_cosvar}
we see that the same relation enters into the derivation of the RLF, and
only the net effect is important for our final results.

\subsection{Spectral modification}
\subsubsection{Epoch-dependent modification factors}

We take a proton spectrum $F_g(E_g)$ injected at some epoch $z_g$, i.e.~the
number of particles in an energy interval d$E_g$ about the injection energy
$E_g$ emitted over a time d$t_g$, and consider its evolution in time,
respectively cosmological epoch $z$.  In the case of continuous energy
evolution, the differential particle number $F(E)\d E\d t$ must be constant
in time -- $F(E)$ at any epoch $z$ is governed by the evolution of d$E$ and
d$t$ alone. Hence, the observable flux $j_s$ at an energy $E_0$, contributed
by a source at $z_g$, is connected to the injection at an energy $E_g =
E_p(E_0,z_g)$; to compare observed flux injection spectrum {\em at the same
energy} $E_0$ we introduce the modification factor\footnote{We adopt this
term from BG88 in its general definition},
	\beqar
	\label{mod}
	M(E_0,z_g) &\equiv& \frac{4\pi d_L^2 j_s(E_0)}{F_g(E_0)} \\
	&=&  	(1+z_g)\;\frac{F_g(E_g)}{F_g(E_0)}\;
		\frac{\partial \Ep}{\partial E_0}(E_0,z_g)\quad.\nonumber
	\eeqar
Here $d_L$ is the luminosity distance of the source, defined as the ratio of
absolute and apparent bolometric luminosity, which can be expressed as a
function of $z$ in a given cosmology (see, e.g., Weinberg 1972). In principle,
$M(E_0,z)$ compares the energy distribution of a number of particles injected
at epoch $z$ to that of the same particles at present, no matter how they are
distributed in space. On the other hand, the connection to a measured flux by
introducing $d_L$ is only correct if the particles propagate rectilinear.
For charged particles, this implies a negligible strength of the overall
magnetic field, a probably unrealistic restriction that must be considered
when \eq{mod} is used to calculate spectra contributed by single sources
with an observed redshift $z_g$.

\Fig{mod} shows modification factors for spectra injected at various epochs,
for injection spectra $F_g(E) \propto E^{-2}$ with and without an exponential
cutoff at $300\,$EeV; the effect of the injection cutoff is of
particular interest for the comparison of the results of BG88 and HS85.  The
Greisen-cutoff and the bumps of piled up UHE particles are essential features
of the modification, as already reported in the references above.  However,
we should stress that the continuous energy loss aproximation tends to
overestimate the height of the bumps, and the sharp peaks at the edge
of the Greisen-cutoff especially for the injection at high redshifts might be
totally artificial.
\begin{figure}[t]
\vspace{9.5cm}
\caption[]{\small\label{fig_mod} Modification factors for proton spectra
($
\alpha_p
= 2.0$) injected at epochs $z$ as indicated. Dashed lines correspond to an
exponential injection cutoff at 300\,EeV ($h_0=0.75$, $q_0 = 1/2$)}
\end{figure}

\subsubsection{Integrated spectra and the influence of source evolution}

Using the modification factor defined in \eq{mod}, the total CR flux
contributed by \FRII galaxies is given by
	\beq\rr
	\label{crspec}
	j_{\rm tot}(E_0) = \frac{1}{4\pi}
		\int\limits_{{\mit z}_{min}}^{{\mit z}_{max}} \d z\;M(E_0,z)\,
		\left[\frac{\widetilde\Psi(E_0,z)}{\tilde d_L^2}\right]\,
		\D{\widetilde V_c}{z}
	\quad,\eeq
with $\Psi(E,z)$, given by \eq{psi}, including our fudge factor. We applied
\eq{RLF_cosvar}, marking cosmology dependent functions taken for the parameters
for which the RLF is given with a tilde; the only function where {\em our}
choice of cosmology appears is $M(E_0,z_g)$. The redshift limits of
integration $z_{min}$ and $z_{max}$ are given by the redshift of the nearest
\FRII galaxie and the epoch of radio galaxy formation, respectively. In the
following we use $z_{min} = 0.3$ (redshift of 3C98), while $z_{max}$ is
clearly model dependent and is given by Peacock as $z_{max} = 10$ for RLF~1 to
3 and $z_{max} = 5.0$ for RLF~4. However, this uncertainty only enters at low
energies, while $z_{min}$ determines decisively the energy and sharpness of
the Greisen-cutoff in the integrated spectrum -- simply integrating from $z=0$
would yield a totally different spectrum at $100\,$EeV, since it assumes non
existent local sources for which the Greisen-cutoff is absent.

\subsection{Particle propagation through space}

In all the discussion presented above we only talked about the evolution of
proton spectra in time; the spatial propagation was always assumed to be
rectilinear. To apply our method to the calculation of proton spectra from
single sources with a clearly defined spatial distance, this assumption is
crucial; if there is any appreciable magnetic field in extragalactic space,
the particles propagate by a combination of drift, convection and diffusion,
so that our results are no longer valid. In \sect5 we discuss the constraints
to the magnetic field for which our results remain useful to a good
approximation.  On the other hand, in our integrated spectra we assume the
source function to be homogeneous in space (this is a basic assumption in
Friedmann cosmology), neglecting gradients in the particle density that govern
diffusion. Obviously, we can still rely in our results for the total CR
spectrum if we can assume that the universe is nearly homogeneously filled
with cosmic rays, which implies that the particles must be able to propagate
over distances at least of the order of the mean distance between \FRII
galaxies (about $100\,$Mpc). Hence, we have to consider how far an average
particle can depart from its source within its lifetime, which is given by
\fig{tau} for a given particle energy at observation.

In fact, nothing is known about the strength and the structure of the
extragalactic magnetic field; usually, its strength is guessed at some
nano-Gauss.  So far, we can only try to discuss constraints within our model
under simplified assumptions. Considering first a random walk with a mean free
path equal to the gyration radius of the particle (Bohm diffusion), we get
	\beq\rn
	\label{diff1}
	d_{max} \approx 30\;
 		\left(\frac{\tau_{17} E_{18}}{B_{-9}}\right)^{\TS\frac{1}{2}}
		\;\rm Mpc\quad.
	\eeq
Here, we take $\tau$ in units of $10^{17}\,$sec (see \fig{tau}), $E$ in EeV
and $B$ in nG. In the energy range between $3$ and $100\,$EeV, we get
$d_{max}\approx 100\,$Mpc in fields of nano-Gauss strength. To introduce a
determined turbulence spectrum in our calculation would be too speculative. We
rather want to discuss as a second possibility a magnetic bubble model,
where we have only one dominant turbulence scale $d_c$ (Ginzburg \&
Syrovatskii 1964). This scale could be given by the distance between galaxies
(typical 1\,Mpc) as well as by the large scale structure of the universe ($\sim
100\,$Mpc). Particles with gyration radii $r_g\ll d_c$ will spiral through the
bubbles with a mean velocity $c/2$ and propagate diffusively with a mean free
path
$d_c$. In this case we simply have $d_{max} = \sqrt{c
\tau d_c/2}$ and
therefore
$d_{max} \gsim 100\,$Mpc if $d_c \gsim 10\,$Mpc independent on particle
energy. In nano-Gauss fields varying over supercluster scales we can apply
this situation to all cosmic ray protons, since $r_g\approx 100\,$Mpc for $E
= 100\,$EeV -- hence, a homogenous filling of the universe is provided for all
particle energies. On the other hand, if $r_g \gsim d_c$, in every bubble a
particle is deflected by a mean angle $\Delta\phi = d_c/r_g$, and we get
\beq\rn
	\label{diff2}
	d_{max} = 30\;\left(\frac{E_{18}}{B_{-9}}\right)
		\left[\frac{\tau_{17}}{d_{c6}}\right]^{\TS\frac{1}{2}}
		\;\rm Mpc\quad.
	\eeq
Assuming nano-Gauss fields varying over scales of $1\,$Mpc, \eq{diff2} can be
applied to particles with energies above $1\,$EeV; therefore, $d_{max} \gsim
100\,$Mpc is provided for $E > 3\,$EeV; particles of highest energies can even
be treated to propagate almost in straight lines. In conclusion, in all
models discussed the spatial distribution of cosmic ray protons above $3\,$EeV
can be treated as homogeneous, if only the strength of the extragalactic
magnetic field is of order $10^{-9}\,$Gauss or smaller (see also Rachen \&
Biermann 1992b).

\section{Results}

In the following figures we compare the results of \eq{crspec} to the UHE-CR
data points of four air shower experiments: the large ground arrays of Akeno,
Yakusk and Havarah Park, and the air fluorescence detector ``Fly's Eye'' in
Utah. Here, we make use of our fudge-factor as a free quantity, which is
determined by normalizing our calculated spectra to the measured flux at
$10\,$EeV. Using \eq{fudge}, we can derive the required $k_p$ for each case,
assuming that all other errors in our estimates are of order 1. We
investigated the influence of all parameters not exactly known in our model;
the cosmological parameters $H_0$ and $q_0$, the source evolution model, the
mean injection spectral index $\bar \alpha_p$ and the mean cutoff energy
$E_c$. However, we only present figures if they contain useful information,
otherwise we just give a qualitative description of the results. Furthermore,
we present figures which show the contribution of nearby single sources, but
without comparing it to any data, since the interpretation of these curves
depends on the details of particle propagation.

\begin{figure}[p]
\vspace{9.1cm}
\caption[]{\small\label{fig_cosvar} Ultra-high energy cosmic ray spectrum;
dependence on cosmology ($H_0 = 100\,h_0\,\rm km\,sec^{-1}\,Mpc^{-1}$)}
\end{figure}
\begin{figure}[p]
\vspace{9.1cm}
\caption[]{\small\label{fig_cutvar} Ultra-high energy cosmic ray spectrum;
dependence on injection cutoff energy $E_c$}
\end{figure}

\subsection{Effect of cosmology and source evolution}

As expected from \fig{evol}, our calculations confirmed that the choice of the
radio galaxy evolution function has no considerable effect on the final
results, except for energies slightly below $1\,$EeV. This expresses the good
knowledge of the distribution and evolution of bright objects as \FRII radio
galaxies up to moderate redshifts, while very distant sources can only
contribute at relative low energies.  In the following we only use Peacock's
RLF~2, since it appears to us to be founded on the most reasonable boundary
conditions. \Fig{cosvar} shows the influence of Hubble's constant on the
spectrum. It turns out that effect of $q_0$, i.e. the geometry of the universe,
is negligible; this is not very surprising, since we already
noticed that this large scale geometry acts on both, the
connection between radio source counts (observed) and the corresponding
evolution model, and on the cosmic ray spectrum derived from it. In
contrary, Hubble's constant acts also on smaller scales, since it is the
overall factor between redshifts and real distance. Hence, its influence on
our results is well noticable: A higher $H_0$ shifts the position
of the Greisen cutuff to somewhat higher energies, but a small $H_0$ improves
the fit to lower energy data points about $1\,$EeV. However, the main effect
of $H_0$ introduces in our fudge factor; a higher $H_0$ implies a lower total
energy in UHE protons necessary to provide the measured flux.

\subsection{Dependence on the injection spectrum}

Much more interesting is the influence of the mean injection spectral index of
the cosmic rays. There is not very much change in the spectral shape when we
change $\bar\alpha_p$ from $1.9$ over $2.0$ to $2.1$ (see Rachen \& Biermann
1992a for a figure), but the corresponding fudge factor $f$ rises from $0.3$
over $4.8$ up to $140$ for $\gamma_c = 10^{11}$. Since for the upper limit of
plausibility $k_p \approx 1000$ we have $f \approx 30$, injection spectra
considerably steeper than $2.0$ (on average) must definitely be excluded. This
supports the expectation of refined acceleration models for the power law
spectra being flatter than in the canonical theory rather than steeper.
However, using the canonical value $\alpha_p = 2$ we have also fudge factors
which are consistent with $k_p < 100$, so that we do not rely on a
modification of the acceleration model.

\Fig{cutvar} shows the influence of the mean cutoff energy. We see that the
spectra for $\bar \gamma_c \ge 10^{11}$ rather fit to the data above
$10\,$EeV, while $\bar\gamma_c \approx 3\cdot 10^{10}$ allows a fit of the
cosmic ray spectrum between $1\,$EeV and $10\,$EeV. We also see that the
Greisen-cutoff gives the real high energy limit to our spectra, while $E_c$
only influences the shape at the high energy end. However, a steepening of the
injection spectrum slightly below the cutoff energy might improve our fit, at
the expense of a higher fudge factor. Here we have to remember that our
assumption of an exponential cutoff at one defined mean energy is rather crude
and disregards the wide spread of the physical conditions in the different hot
spots as well as the overlay of different cutoff effects at probably different
energies in every source. Therefore, the real shape of the averaged injection
spectrum might be badly approximated.

\subsection{Contribution of single sources}

In \sect4 we determined the confinement volume for cosmic ray protons
observed at at certain energy $E_0$. For energies above $30\,$EeV
this volume gets so small that only about a dozen \FRII galaxies contribute,
and we must ask if this should imply an anisotropy in the distribution of CR
events. \Fig{singal} shows the {\em total} contribution of three prominent
\FRII galaxies in our neighborhood, assuming $E_c \approx 300\,$EeV; note that
the total flux is plotted here and not the flux per unit solid angle. The
kinetic jet power $P^{(0)}_{jet}$ of 3C33 is taken from RS91
($9\cdot10^{37}\,$W), that of 3C111 and Pic~A ($7\cdot10^{37}\,$W and $1\cdot
10^{37}\,$W, respectively) is calculated out of the single hot spot results
from M89, assuming that the same power is contributed in both hot spots of the
radio galaxy. Furthermore, we assumed $k_p = 30$ consistent with the fudge
factors found for the total spectrum. All single source spectra are valid {\em
only for straight line propagation}, i.e. a negligible strength of the
intergalactic magnetic field, in contrast to the integrated spectra where we
can argue that the propagation of the individual particles can be disregarded
if a homogeneous filling of the universe with cosmic rays is provided (see
\sect{3.4}).
\begin{figure}[p]
\vspace{9.1cm}
\caption[]{\small\label{fig_singal} Total proton flux of nearby
\FRII galaxies; the
spectral indices are deduced from their synchrotron spectra}
\end{figure}

We see that Pic~A is expected to produce the main part of all observed CR
particles at about $80\,$EeV, but this is strongly connected to its
flatter particle spectrum (inferred from the synchrotron emission); setting
the spectral indices for all sources to $2.0$, Pic~A is no longer dominant
compared to more distant sources like 3C33 and 3C111. Additionally,
\fig{pica} shows how the strong bump shown in \fig{singal} is reduced with
decreasing $\gamma_c$, what is expected since the number of particles that
pile up in the bump is deminished.
\begin{figure}[p]
\vspace{9.1cm}
\caption[]{\small
\label{fig_pica} Dependence of total proton spectrum expected
from
Pictor A on the injection cutoff}
\end{figure}

The question is now: If the power law index of Pic~A is really as flat as
inferred from the synchrotron spectrum (i.e., if the spectral index is equal
for electrons and protons), can we expect that the cosmic ray events at this
energy are peaked about the position of Pic~A? Clearly, in the case of strong
diffusion in a tangled magnetic field our method fails to make any statements
about the contribution of single sources, since we would have to solve a
spatial diffusion equation.  But if the diffusion is so weak that the particles
only scatter a little around the straight line connection between us and the
source, we can treat this movement as a random walk in angles (Ginzburg \&
Syrovatskii 1964). Since the particles mainly move in a definite direction,
\eq{mod} and Figs.~(\ref{fig_singal}) and (\ref{fig_pica}) remain valid
in good approximation. Considering the magnetic bubble model with a typical
bubble dimension $d_c$ in Mpc and a field strength $B_{-9}$ in nG, a particle
from
a source at distance $d_S$ with a gyration radius $r_g\gsim d_c$ is deflected
by a mean angle
	\beq
	\label{diffs}
	\phi \approx \frac{B_{-9}\sqrt{d_S d_c}}{E_{18}} \quad,
	\eeq
In the case of nano-Gauss fields and cell dimensions of about $1\,$Mpc, most
particles from Pic~A ($d_S \approx 100/h_0\,$Mpc) with energies above
$30\,$EeV will reach us from within one steradian about the position of the
source; hence, if Pic~A is really the dominant source of particles of highest
energies, there should be an anisotropy connected with its position. In
contrast, a magnetic field mainly varying over supercluster scales would
rather imply an anisotropy connected to the local field direction.  Large
scale fields should be of order pico-Gauss to maintain the observability of
local \FRII galaxies as cosmic ray sources. Additionally, the galactic disk
may also influence the distribution, in particular it might obscure \FRII
sources near the direction of the galactic center.

\section{Discussion}

\subsection{Injection spectrum}

We have seen that the exact form of the averaged injection spectrum plays an
important role for the shape of the predicted UHE-CR spectrum. From
\fig{cutvar} we see that a spectrum that is steepened in the EeV energy regime
due to cutoff effects may even fit better to the data, especially below
$10\,$EeV, than an unmodified spectrum.  Clearly, to increase the overall
spectral index $\alpha_p$ would be fatal, since we cannot provide the observed
flux in this case. On the other hand, a sharp cutoff below $100\,$EeV (like
that one assumed in \fig{cutvar}) prevents the explanation of the data above
$10\,$EeV. To obtain a more reliable injection spectrum the exact (or at least
approximated) solutions of the transport equation in a hot spot like magnetic
topology must be found. Moreover, the distribution of the typical hot spot
properties about their mean values must be taken into account. Unfortunately,
the theoretical and observational basis for that purpose is rather poor yet. In
particular, the spread in the cosmic ray data offered by the various
experiments is so large that a comparison with calculated spectra is
difficult.

\subsection{Anisotropy}

At present, there is no reliable evidence for any anisotropy in the cosmic ray
events above $1\,$EeV; but it must be considered that it is rather difficult
to derive information about anisotropy in events that occur with a rate of
only a few per year, taking into account all experimental selection effects
(see, e.g., Sokolsky 1989).  Statements about cosmic ray anisotropy depend in
any case on the location of the sources respectively assumed: the galactic
center, the galactic disk in general or the Virgo cluster.  Thus, our model
does nothing else than to present new ``seeds'' to look for anisotropy.
Recently, Chi \ea.~(1991) published a map of groups of UHE-CR events and
proclaimed a good correlation with the galactic disk, but it is interesting to
note that the correlation to the positions of nearby \FRII galaxies appears to
be just as good. If this correlation can be confirmed, it would imply severe
restrictions to the intergalactic magnetic field. On the other hand, if any
considerable anisotropy of UHE-CR events can be ruled out, it would be no
definite falsification of our model; there must be an independent proof that
the extragalactic magnetic field is weak enough to allow an almost rectilinear
propagation of UHE protons.

\subsection{Heavy nuclei}

{}From the cosmic abundances of elements, it is reasonable to assume that
protons, i.e. hydrogen nuclei, make the main part of all cosmic rays; $\alpha$
particles should be suppressed by one, heavier nuclei by two orders of
magnitude. On the other hand, heavy nuclei might get important in energy
regions where protons are supressed by the limits of acceleration and
cosmological transport. There are several advantages and disadvantages of
heavy nuclei with charge $Z$ and mass $A$ with respect to the explanation of
the maximum energy of cosmic rays: (i) their gyration radii are smaller by a
factor $Z$, thus the diffusive particle losses in the acceleration process are
of less importance, but on the other hand it causes a stronger diffusion in
the extragalactic magnetic field; (ii) the energy limits for the production of
secondary particles at the MBR are increased by a factor of $A$, but
additionally heavy nuclei can be destroyed by photo-disintegration at energies
below the photomeson-prodution threshold. The energy losses for heavy nuclei
are generally higher than for protons with the same {\em Lorentz factor}, but
the confinement volume at the same {\em energy} is enlarged and as a result
the Greisen-cutoff for $^{56}$Fe is shifted to energies above $100\,$EeV
(Puget \ea.~1975). Although heavy nuclei might play an important role at
energies near the Greisen-cutoff, we cannot agree with the opinion of Ip \&
Axford (1991) that {\em all} cosmic rays above $10\,$EeV should be heavy
nuclei rather than protons. It is obvious from our calculations that
unreasonable high fudge factors are required to establish the observed CR flux
if the most abundant component is removed.

\subsection{Secondary backgrounds}

Extragalactic UHE protons and other nuclei lose a lot of energy on their way
through the universe by interactions with the MBR, which is channelled into
secondary particles.  Thus, a considerable diffuse extragalactic neutrino and
$\gamma$ background should result from this reactions.  Neutrinos originate
only in photomeson reactions and can easily propagate over cosmological
distances. Therefore, they could give information about the early (unmodified)
UHE-CR population above the threshold for photomeson production, if their flux
can compete with the atmospheric flux and the neutrino flux expected directly
from astrophysical sources (Mannheim \ea.~1992). On the other hand,
$\gamma$-photons originate mainly from electromagnetic cascades, initiated by
$\gamma p$ pair production or by UHE photons itself ($\gamma\gamma\to e^+e^-$).
In the UHE regime, the interaction length of photons against $\gamma\gamma$
reactions with ambient backgronds is so short that they only give an image of
the present particle spectrum, their contribution is expected to be about
$10\%$ (e.g. Halzen 1991). Below $0.1\,$PeV, however, the $\gamma$-spectrum is
influenced by the cosmic ray population at earlier epochs and may reach the
observational limits, hence the calculation of the implied $\gamma$--spectrum
could be a crucial test of our model.

\subsection{Other sources and the overall CR spectrum}

It is straightforward to apply our model to other possible sources of cosmic
rays to give a full description of the extragalactic contribution to the CR
spectrum. A similar model is already proposed by Protheroe \& Szabo (1992),
which investigate protons that escape via spinflip (i.e., as neutrons) from
the cores of quasars -- they predict a strong contribution in the PeV--EeV
regime. However, cosmic rays in this energy regime may also be explained as of
galactic origin (Biermann 1993). A test would be an observation of
the proton component only, since the extragalactic model predicts a large
proton flux, while the galactic model requires a strong contribution of heavy
nuclei. In this way extragalactic predictions could even be tested at
energies where their flux does not dominate over the galactic contribution.

A simple extension of the present model is the inclusion of FR-I galaxies;
they show jets and knots, where also shock waves can be expected and where
particles should be accelerated by the same method as in \FRII hot spots.
However, FR-I knots are much smaller than \FRII hot spots, and they are
located inside the galaxy so that adiabatic losses of the energetic particles
on their way out must be taken into account. Certainly, the transport
calculations of BG88 can be applied to any kind of galaxy population if only
their evolution function is known and there is some model of their cosmic ray
injection. It is easy to show, however, that normal galaxies cannot inject
particles with a simple $E^{-2}$ spectrum, because this would give far too
high fluxes at high energies (Biermann 1991), but it is worth to investigate
their expected contribution within a more detailed model.

\section{Conclusion and outlook}

We have presented a model of the origin of ultra-high energy cosmic rays above
about $1\,$EeV, assuming hot spots in \FRII radio galaxies as their sources
and diffusive shock acceleration as the mechanism how they attain their
energy. We have shown that the observed fluxes and energies of the cosmic ray
particles can be explained by this model, using the canonical first order
Fermi theory and including all present knowledge about source evolution and
cosmic ray transport over cosmological distances.  We also have seen that the
injection spectra must be on average $E^{-2}$ or slightly flatter to provide
the observed flux, what agrees with the predictions of enhanced acceleration
models, but still includes that of the canonical theory. We expect the
injection spectrum to be steepened at highest energies, which would improve
the fit to the data. Because of the large distance of the nearest \FRII galaxy
the Greisen-cutoff appears necessarily in our resulting spectra. At energies
below $1\,$EeV, the spectrum depends a little on the evolution model of radio
galaxies, but in general it is obvious that other sources must dominate here.
We disregarded heavy nuclei in our consideration, but our results make clear
that they can only be important at energies near or above the proton cutoff at
about $80\,$EeV; the main flux can only be provided by protons, if we assume
cosmic abundances of elements in jets and hot spots. At last, we found it
possible that an observable anisotropy may be caused by the rareness of local
sources that can solely contribute at highest energies.

In following papers we will take more effort on getting a reliable mean
injection spectra, including cutoff effects at least in some reasonable
approximation and also including the spread of the physical conditions in hot
spots. However, present data on hot spots are rather sparse and we can only
hope that the observational material can be enlarged in the future.  To get a
better foundation for comparison with the cosmic ray data, we will also
present a method of averaging the present data over various experiments,
taking into account the possible sytematic errors in energy calibration.
Later, our model should be enhanced by considering heavy nuclei and secondary
photon and neutrino backgrounds. To provide this, the model must be able to
deal with particle transitions, i.e., the injection of lighter nuclei or
secondaries from interactions of heavier ones.  Our predictions concerning a
possible anisotropy could be improved by using a correct diffusion
approximation, but this may be deferred until we get either better
observational evidence for such an anisotropy or we have more concrete
estimates on the extragalactic diffusion coefficient.

\subsection*{Acknowledgements}
We have profited for this paper especially from discussions
with Drs.~K.~Mannheim and T.~Stanev. The up-to-date cosmic ray data sample we
used was kindly provided by Drs.~P.~Zhang and P.~Sokolsky.

\subsection*{References}
\frenchspacing
\begin{small}
\begin{list}{}
{\labelwidth=\parindent \leftmargin=\parindent \labelsep0pt
\parsep0pt \itemsep0pt plus 2pt}
\item[Ballard K.R., Heavens A.F.], 1992, MNRAS 259, 89
\item[Berezinsky V.S., Grigor'eva S.I.], 1988, A\&A 199, 1
({\bf BG88})
\item[Biermann P.L.], 1990, in: Fazio G.G., Silberberg R.
(eds.), {\em Currents in Astrophysics and Cosmology}, \\
Cambridge University Press, publication expected in June 1993
\item[Biermann P.L.], 1991, in: Nagano \ea. 1991, p. 301
\item[Biermann P.L.], 1993, A\&A (in press)
\item[Biermann P.L., Strittmatter P.A.], 1987, ApJ 322, 643
({\bf BS87})
\item[Blumenthal G.R.], 1970, Phys. Rev. D 1, 1596
\item[Chi X. \ea.], 1991, in: Nagano \ea. 1991, p. 140
\item[Dickel J.R., Breugel W.J.M.van, Strom R.G.], 1991, AJ 101, 2151
\item[Drury L.O'C.], 1983, Rep. Prog. Phys. 46, 973
\item[Ellison D.C.], 1991, in: Nagano \ea. 1991, p. 281
\item[Fanaroff B.L., Riley J.M.], 1974, MNRAS 167, 31P
\item[Forman M.A., Webb G.M.], 1985, in: Stone R.G., Tsuratani
B.T. (eds.), {\em Collisionless Shocks in the Heliosphere: A Tutorial Review},
AGU Monograph Vol. 34; AGU, Washington D.C., p.~91
\item[Ginzburg V.L., Syrovatskii S.I.], 1964,
{\em The Origin of Cosmic Rays}, Pergamon Press, London
\item[Greisen K.], 1966, Phys. Rev. Lett. 16, 748
\item[Gush H. \ea.], 1990, Phys. Rev. Lett. 65, 537
\item[Halzen F.], 1991, in: Nagano \ea. 1991, p. 91
\item[Heavens A.F.], 1989, in: Meisenheimer \& R\"oser 1989, p. 247
\item[Hill C.T., Schramm D.N.], 1985, Phys. Rev. D 31, 564
({\bf HS85})
\item[Hillas A.M.], 1968, Can. J. Phys. 46, S623.
\item[Ip W.-H., Axford W.I.], 1991, in: Nagano \ea. 1991, p. 273
\item[Jokipii J.R.], 1987, ApJ 313, 842
\item[Kirk J.G., Heavens A.F.], 1989, MNRAS 239, 995
\item[Kirk J.G., Schneider P.], 1987, ApJ 315, 425
\item[Kr\"ulls W.M.], 1992, A\&A 260, 49
\item[Mannheim K.], 1993, A\&A, in press
\item[Mannheim K., Biermann P.L.], 1989, A\&A 221, 211
\item[Mannheim K., Biermann P.L.], 1992, A\&A 253, L21
\item[Mannheim K., Kr\"ulls W.M., Biermann P.L.], 1991,
 A\&A 251, 723
\item[Mannheim K., Stanev T., Biermann P.L.], 1992, A\&A 260, L1
\item[Mather J.C. \ea.], 1990, ApJ 354, L37
\item[Meisenheimer K., R\"oser H.J. (eds)], 1989, {\em Hot
Spots in Extragalactic Radio Sources};\\ Proc. Ringberg Castle, 1988;
Springer, Heidelberg
\item[Meisenheimer K. \ea.], 1989, A\&A 219, 63 ({\bf M89})
\item[Nagano M. \ea. (eds)], 1991, {\em Astrophysical Aspects of
the Most Energetic Cosmic Rays};\\ Proc.~ICRR, Kofu, 1990;
World Scientific, Singapore
\item[Pacholczyk A.G.], 1970, {\em Radio Astrophysics},
Freeman and Company, San Francisco, CA
\item[ Particle Data Group], 1988, Phys. Lett. B 204
\item[Peacock J.A., Gull S.F.], 1981, MNRAS 196, 611
\item[Peacock, J.A.], 1985, MNRAS 217, 601 ({\bf P85})
\item[Perley R.A.], 1989, in: Meisenheimer \& R\"oser 1991, p. 1
\item[Protheroe R.J., Szabo A.P.], 1992, Phys. Rev. Lett. 69, 2885
\item[Padovani P., Urry C.M.], 1992, ApJ 387, 449
\item[Puget J.L., Stecker F.W., Bredekamp J.H.], 1975,
ApJ, 205, 638
\item[Rachen J.P., Biermann P.L.], 1992a, in: Meisenheimer K.,
R\"oser H.J. (eds), {\em Jets in Extragalactic Radio Sources}; Proc.
Ringberg Castle, 1991; Springer (in press).
\item[Rachen J.P., Biermann P.L.], 1992b, in: Zank G.P., Gaisser
T.K., {\em Particle Acceleration in Cosmic Plasmas};
Proc.~Bartol, Newark~DE, 1991; AIP conf. No. 264, p. 393
\item[Rawlings S., Saunders R.], 1991, Nature 349, 138
({\bf RS91})
\item[Sokolsky P.], 1989, {\em Introd. to Ultrahigh Energy
Cosmic Ray Physics}, Addison Wesley, Redwood City, CA
\item[Stecker F.W.], 1968, Phys. Rev. Lett 21, 1016
\item[Stecker F.W.], 1989, Nature 342, 401
\item[Takahara F., Terasawa T.], 1991, in: Nagano \ea. 1991, p. 291
\item[Weinberg S.], 1972, {\em Gravitation and Cosmology},
Wiley, New York
\item[Windhorst R.], 1984, Ph.D.~thesis, Sterrewacht Leiden,
Leiden, Netherlands
\item[Zatsepin G.T., Kuzmin V.A.], 1966, JETPh Lett. 4, 78
\end{list}
\end{small}

\end{document}